\documentclass[11pt,a4paper]{article}

 \usepackage{amsmath}
\usepackage{amssymb}
\usepackage{amsthm}
 \usepackage{epsfig} 
 \usepackage{graphics} 

\begin{document}
\bibliographystyle{plain}

\pagenumbering{arabic}
\setcounter{page}{1}

\title{Schizophrenia -- a parameters' game?}
\author{Anca R\v{a}dulescu, Applied Mathematics, University of Colorado at Boulder\\ UCB 526, Boulder, CO 80309-0526, \emph{radulesc@colorado.edu}}
\maketitle

       \begin{align*}
       &\text{\hspace{5cm} {\bf Motto:} \small{``Ivan concentrated his attention on the cat and saw how the}}\\ 
       &\text{\hspace{6cm} \small{strange animal walked up to the platform of an A tram waiting}}\\
       &\text{\hspace{6cm} \small{at a stop, cheekily pushed off a screaming woman, grasped the}}\\
       &\text{\hspace{6cm} \small{handrail and offered the conductress a ten-kopeck piece.''}}\\
       &\\
       &\text{\hspace{8cm} M. Bulgakov -- ``The Master and Margarita''}
       \end{align*}	

\vspace{10mm}

\small{Schizophrenia is a severe, currently incurable, relatively common mental condition. Its symptoms are complex and widespread. It structurally and functionally affects cortical and subcortical regions involved in cognitive, emotional and motivational aspects of behavior. Its cause is unknown, its diagnosis is based on statistical behavior and its treatment is elusive. 

 Our paradigm addresses the complexity of schizophrenic symptoms. Building upon recent neural vulnerability and limbic dysregulation hypotheses, it offers a mathematical model for the evolution of the limbic system under perturbation. Dependence on parameters and the concept of ``bifurcation" could be the key to understanding the threshold between ``normality" and ``disease".}

\section{Stress and mental illness}

In the wild, living beings survive by responding to perceived
threats with adaptive and appropriate changes in their behaviors and
physiological states. Besides the species-specific factors, the
nature of these responses depends on the external environment,
but also on the internal physiological and emotional
conditions. Unfortunately, the neuroendocrine
mechanisms that control stress responses based on these environments
are poorly understood for most animals~\cite{Lowry}, in particular
for humans.

Altogether, we are aware of a few physiological
mechanisms that allow stressful events to affect behavior. The
stress response impacts on autonomic functions, such as respiration
and heart rate variabiltiy~\cite{Lucini}. It also alters, through  the
hypothalamic-pituitary-adrenal axis, the secretion of
corticotropin-releasing factor, the hormone that coordinates
behavior, autonomic and hormonal function in response to
stress~\cite{Aguilera}. It subsequently affects the production of
multiple other hormones, such as adrenocorticotropin hormone and
cortisol.

It has been generally understood for years that, via its autonomic
effects, sustained stress can severely affect health~\cite{Elliott}, contributing to a variety of conditions, among which heart disease, diabetes, growth retardation~\cite{Sapolski}, decresed immunity~\cite{Manuk} and various eating and digestive disorders~\cite{Jones}. Through similarly complex mechanisms, stress is also believed to lead to a number of psychiatric disorders, including depression, posttraumatic stress disorder, Alzheimer's disease, and other anxiety disorders~\cite{Kim}. Interestingly,
whether one is male or female is one of the most important health
determinants. This has been partly explained exacly by the
underlying sex differences in the physiological response to stress,
specifically by the fact that estrogen exposure attenuates
sympathoadrenal responsiveness. It has been hypothesised that these
differences have been driven in females by strong evolutionary
pressure, in the need to protect the fetus from the adverse effects
of maternal stress responses, in particular excess glucocorticoid
exposure~\cite{Kaj}. As psychiatric stress-associated pathologies
will be the object of our present study, we will come back to the
idea of neural exposure to cortisol in a later section.

In relation to behavior and mental illness, stress has been noticed
to impair heathy individuals and psychiatric patients in sometimes
surprisingly similar ways. Consider schizophrenia, for example, wich
will be the focus of this paper: a severe mental
disorder with a heterogeneous set of symptoms including paranoia,
hallucinations, delusional beliefs, thought disorder, emotional
flattening and social withdrawal. The illness is relatively common,
affecting 1.1$\%$ of the population, or over 2.2 million people
every year in the United States alone, and has a devastating impact
on social functioning. Studies over the past two decades have
established that it is a chronic~\cite{1}~\cite{2} and
neurodegenerative~\cite{3}~\cite{4}~\cite{5} disease, structurally
and functionally affecting various cortical and subcortical regions
involved in cognitive, emotional and motivational aspects of
behavior~\cite{6}~\cite{7}~\cite{8}~\cite{9}. 

Many cognitive abnormalities~\cite{68}~\cite{69}~\cite{70} associated with this
illness, such as delusions and hallucinations~\cite{Hirsch},
impaired memory~\cite{61}, lowered sensory gating and selective
attention~\cite{59}~\cite{57} are also induced in healthy adults
under acute emotional stress~\cite{Ghi}~\cite{Silva}~\cite{Brugger}.
So the question raises itself: Can stress single-handedly cause a
mental illness just through its perseverance? And if not, then what
is the intrinsic detail that makes the dramatic difference between
normality and pathology?

Historically, theories have ranged over a large scale, and
psychiatry hasn't lacked controversies. Some attributed 
schizophrenic symptoms solely to prenatal stress, or to 
social and other environmental factors during the patient's
childhood and teenage years~\cite{Jarvis}. Others made them entirely the
responsibility of genetic programming~\cite{Bearden}~\cite{Will}.

Unfortunately, despite intensive research, our knowledge of the
underpinnings of schizophrenia is now about as relevant as
ever. At present, there is no generically sustainable treatment for
most serious psychiatric illnesses. Progress has been made in
understanding some effects and side-effects of medication, but the
drugs that are being used may only treat the effects of the disease
rather than its cause. In fact, the main unanswered questions in
current psychiatry concern diagnosis of mental illness as much as
its treatment. Consensus diagnoses are revised
periodically~\cite{DSM} based upon observed behavior rather than
actual causes (etiology), which are usually still unknown. Most
psychiatric conditions come with complex, widespread symptoms,
accompanied by seemingly unrelated cognitive abnormalities and
psychosocial consequences. The severity of symptoms cannot be stated
in reproducible terms, and is therefore left to clinical
interpretation, which varies among psychiatrists. This increases the
potential of misdiagnosis, together with the fact that psychotic or
very young patients are often neither cooperative nor articulate
enough to describe their symptoms.

This is true in particular for schizophrenia. Recently, an idea which 
gained ground, under diverse forms, has been
that the etiology of schizophrenia is based on neural vulnerability
and degeneration. The theory (first conceptualized in the mid 19th
century) attributes mental disorders to a hereditary predisposition
that reduces the individual psychological threshold towards
stimuli~\cite{Stamm}, to the point where even minor daily stresses
will directly trigger psychotic experiences~\cite{Myin}. It has been
observed that this ``vulnerability" (or lack of inhibition in the
threat detection mechanism~\cite{HBM}) manifests itself as an overt
illness only under the impact of stress factors~\cite{Ventura}, so
that schizophrenic disturbances eventually result as an overlap of
environmental stress onto the individual's premorbid personality
component.

In support to this hypothesis come the well-known relationship
between stress and first-break psychosis~\cite{Hazlett} or
relapse~\cite{Ventura}, and the sympathetic upturn that occurs prior
to symptom exacerbation. Prospective data
suggest that signs and symptoms (such as elevated autonomic
activation~\cite{Aut} and electrodermal activity levels~\cite{Hazlett}~\cite{28})
prodromal to psychotic first episodes and relapses may be present in
about 60 percent of patients. Also, it has been observed that
stressful life events and highly critical attitudes toward the
patient in the social environment predict relapse~\cite{Nuech}.
On the other hand, research with schizophrenic out-patients has
shown that antipsychotic (neuroleptic) medication reduces relapse
rates. This protective factor may operate partially by raising the
threshold in the face of environmental stressors~\cite{Ventura2}.

Such first outbreak and relapse predictors are currently being used
as clinical indicators for schizophrenia, together with more
traditional ones, such as paranoia, agitation and
sleeplessness~\cite{Hirsch}. The subsequent possibility of
pre-symptomatic treatment~\cite{12}~\cite{13}, among other things,
motivated a more careful investigation of the factors implicated in
producing this ``vulnerability", and the attempt to determine the
conditions that precipitate the plethora of outward clinical
manifestations diagnosed as ``schizophrenia". In this
perspective, several vulnerability models have
been proposed~\cite{Berner}, which we will discuss in some detail
in the next section.

\section{The limbic dysregulation hypothesis}

Recent studies have increasingly correlated vulnerability to
schizophrenia to volume reductions in several limbic areas (amygdala,
hippocampus and prefrontal cortex~\cite{Yui}).
In light of the previous section, this should not appear surprising,
since the limbic system is primarily associated with the regulation of
emotion and arousal, and is also responsible for integrating the
internal and external environments via its wide connections with the
neocortex~\cite{Anand}, as well as with the autonomic~\cite{Amann} and
endocrine~\cite{Mason} systems.

Dopamine and serotonin abnormalities ~\cite{Joyce}~\cite{Yam} in
schizophrenia constitute today the most established and popular etiological
hypothesis (which forms the bases for development of newer
antipsychotics~\cite{Fried}~\cite{Kapur}). However, schizophrenia has many
neurobiological features suggesting an underlying dysregulation of
emotional arousal, including limbic~\cite{Chua}~\cite{39},
endocrine~\cite{Ritsner}~\cite{Sap}~\cite{21}~\cite{22}, and
autonomic~\cite{Aut}~\cite{Zahn}~\cite{MP} abnormalities. It is
possible that the neurotransmitter disfunction may be induced by
hyperarousal~\cite{72}~\cite{73}~\cite{74}, making it a consequence
of dysregulation, rather than its cause.

Over the years, this variety of abnormalities of schizophrenia (and
the relationship among themselves and with hyper/hypo-arousal) have
spurred research interest. Historically, Gruzelier et
al.~\cite{23}~\cite{24} were among the first to hypothesize that
limbic abnormalities in the regulation of arousal may be an
important feature of the disease. Grossberg's model~\cite{33} linked
schizophrenia to pathologies of the amygdala and Hanlon and
Sutherland~\cite{34} -- to prenatal damage to the limbic system.
Williams et. al.~\cite{39} have reported differences between the
amplitude of patient and control limbic and autonomic responses to
different facial expressions. Other studies pointed out the
abnormally high cortisol levels in schizophrenic patients,
~\cite{21}~\cite{22}, and McEwen~\cite{38} has sustainably related 
mental illness to chronic stress and to the corresponding deleterious 
effects of cortisol on the hippocampus and prefrontal cortex to mental illness.

In (1989), Nuechterlein and others~\cite{Nuech}~\cite{27}~\cite{28}
elaborated a ``vulnerability/stress" hypothesis of schizophrenia, in
which a vulnerability to stress (assumed to be due to cognitive
deficits), combined with stressful ``life-events," leads to
first-break or relapse of schizophrenia. Advances in
understanding the neurobiology of the stress cascade led to a
plausible model by which this vulnerability may occur through
neurotoxic effects on the hippocampus that may involve synaptic
remodeling~\cite{Corc}~\cite{Kim}~\cite{Weinberger}.

In this context, schizophrenic symptoms may constitute an end-stage
of a cyclic and neurodegenerative process.  Recent
studies~\cite{Medoff}~\cite{Preston}~\cite{Tamm} support the theory
that the vulnerability to stress in schizophrenia is based on a
pre-existing hippocampal/prefrontal deficit. Impaired
hippocampal/prefrontal function leads to decreased inhibition of the
amygdala, contributing to higher arousal levels, even under minor
stress. Via the connections of the amygdala with the hypothalamus,
the fear reaction triggers autonomic and endocrine effects~\cite{45}
(such as changes in heart rate variability~\cite{MP} and
electrodermal activity~\cite{Hazlett}~\cite{28}, or increased
cortisol levels~\cite{Sap}). Excessive cortisol leads to brain
neurotoxicity~\cite{Weinberger} and further hippocampus
damage~\cite{64}~\cite{Sap}, thus closing the dysregulation vicious
cycle. The delay in schizophrenia's onset (late teens in males and
early 30's in females) is consistent with a vicious cycling process,
in which the neurodegenerative loop would need sufficient time to 
progress to the point where symptoms become
apparent; the already mentioned neuroprotective effect of estrogen
in ameliorating the effects of cortisol on the hippocampus and
prefrontal cortex~\cite{Bao}~\cite{Grant} would slow but not stop
the cycle in females~\cite{Kessler}, accounting for their delayed
onset.

The dynamical analysis in this paper is based on a control system
model described by Sotres-Bayon et al.~\cite{SB}, in which limbic regions define a
negative feedback loop that regulates arousal. The central amygdala
forms the main excitatory component of the arousal
response~\cite{Davis}. The primary inhibitory pathways are the
medial prefrontal cortex~\cite{Baxter}\cite{44}
~\cite{Izq}~\cite{Izq2}~\cite{Phelps}~\cite{Rosen} and the
hippocampus~\cite{Corc}~\cite{SB}. Outputs from the limbic system,
via the hypothalamus, provide inputs for the endocrine and autonomic
nervous systems. In this sense, the model presents a review of the
known possible mechanisms for regulation of arousal. In our context,
the model explains how limbic dysregulation in schizophrenia
could lead to its characteristic behavioral features and could also
cause the endocrine and autonomic abnormalities that so often
accompany the illness.

Section 3 will give a brief review of the known neural pathways that
underlie the limbic connection assumed in our model. Section 4 will
construct and analyze the mathematical model. Section 5 will
interpret the results of the analysis and will discuss the
conclusions in a clinical context.

\section{Connections and pathways}

Over the past decade, significant research has been conducted on the
role of the prefrontal cortex, the hippocampus  and the amygdala in
the fear conditioning and extinction. The predominant view is that
the  amygdala is excitatory and the hippocampus and prefrontal
cortex are inhibitory~\cite{SB}. More precisely, we believe that the
activity of the prefrontal cortex modulates the amygdala fear
reaction to a stressor. In this section, we will describe in some
detail the internal anatomical organization and the pathways between
the regions involved in the stress-reaction. This will provide us
with some background and motivation for our mathematical model,
although the model itself will be much more schematic and will try
to avoid detail.

\begin{enumerate}

\item {\bf Amygdala.} It has been observed, in both human and animal
studies, that damage to the amygdala prevents the acquisition and
expression of fear. It was thereby concluded that the amygdala may
be the underlying site for fear conditioning and extinction.
Amygdala is divided into a few physiologically and functionally
distinct parts: the lateral amygdala (LA), the central amygdala
(CE), the basal nucleus (B) and the intercalated cell mass (ITC).
The current hypothesized mechanism of the fear reaction, in a very
simplified form, is the following: In the absence of stimuli, the 
intra-amygdala connections are suppresing its activation,maintaining it at fairly low levels. When an emotionally potent conditioned stimulus is received, it is transmitted via thalamic pathways to the LA, then to
the CE (either directly or via more complex intra-amygdala
connections). Finally, the CE has output connections to a set of
regions that control specific autonomic, endocrine and behavioral
responses (autonomic and endocrine systems, PFC). The role of B is
still controversial. Although there is anatomical~\cite{Pitkanen} 
and physiological~\cite{Ishikawa} evidence
that there are strong reciprocal projections of B with the
hippocampus and with the mPFC, B lesions seem to have no effect on
fear extinction. It has been suggested that the role of B may be to
integrate information from the LA, hippocampus and mPFC, thus being
a site of contextual contributions to conditioning. As both the
hippocampus and the PFC are belived to be crucial in the dynamics
of schizophrenia, and as contexetual interpretation of threat has
been proved to be impaired in schizophrenic patients, these
interconnections may prove to be of interest to our present
study.

\item {\bf Prefrontal cortex.}

Damage to the prefrontal cortex (PFC) is known to generally induce
emotional and cognitive changes. In fact, these changes seem to be
very finely-tuned and region-specific. The PFC consists of several
functionally distinct subregions, wich include the lateral prefrontal cortex, 
the orbital frontal cortex and the medial prefrontal cortex
(mPFC)~\cite{Muller}~\cite{Seamans}~\cite{Robbins}. The lateral prefrontal cortex is involved in working memory and executive control functions (such as motor control)~\cite{Miller}. The orbitofrontal
cortex is involved in motivation, reward and emotional
decision-making~\cite{Damasio}~\cite{Berns}. The mPFC is itself
divided into a few subregions: anterior cingulate cortex (ACC) and
several more ventral areas (infralimbic, prelimbic). The dorsal part
of the ACC is involved in attention and cognitive control, and the
ventral part in emotional regulation~\cite{Bush}. The functionality
of the other subregions hasn't yet been clearly esablished, but the predominant view is that neural activity in the mPFC regulates the amygdala-mediated fear responses via direct projections to the LA or the ITC, as well as the activity in the hippocampus, via projections to CA1 (see below). Conversely, experimental studies suggest that initiating and sustaining behavior require several types of mPFC modulation, including mPFC self-stimulation~\cite{Mora}~\cite{Ferrer}.

\item {\bf Hippocampus.} 
The hippocampus is critical in episodic memory cosolidation~\cite{Squire} and for aspects of working memory~\cite{Lipska}. Unlike the role of the amygdala and PFC in stress processing, which have been confirmed by a wide variety of studies, the potential contribution of the hippocampus remains relatively unexplored. 

Structurally, MRI studies~\cite{Caet} found decreased hippocampal volumes in depressed patients and correlated the volume loss with the length of the illness. The same volume reduction has been observed in schizotypal disorders~\cite{Dickey}. This is consistent with the hypothesis that hypercortisolism could result in hippocampal neurotoxicity in conditions such as bipolar disorder and schizophrenia. 

However, although cronic stress has been shown to structurally damage the hippocampus, this damage is believed to be restricted to particular subfields~\cite{Sousa}~\cite{McEwen}, which is possibly not sufficient to explain psychotic symptoms. Cerqueira et al.~\cite{Cerq} showed that chronic stress may also impair working memory and behavioral flexibility indirectly, by affecting not the volume or the number of neurons in the hippocampus itself, but rather the synaptic plasticity within CA1~\cite{Kemp} or of the hippocampus-PFC interactions (see the paragraph below on hippocampus-PFC pathways).

\item {\bf Amygdala -- prefrontal cortex}

Different amygdala nuclei are robustly connected with different regions in the mPFC, suggesting that the two are functionally coupled. Several studies have shown that the functional mPFC activity is inversly related to  amygdala activity~\cite{Anand}, and this regulatory interaction is believed to be critical for the organism's ability to adapt to change. Although it has been proposed that mPFC inhibits activity in the amygdala, the mechanisms of this suppression are not yet known. As most mPFC projections to the amygdala are excitatory, it has been proposed that the inhibition occurs by activation of inhibitory neurons within the amygdala~\cite{Rosen}. However, based on experimental evidence, a new study~\cite{Vidal} suggests a more complex, bidirectional modulation of fear, in which PL excites amygdala output (via its projections to B) and IL inhibits amygdala output (through its projections to LA and ITC).

It has been argued that dysfunction of the mPFC-amygdala interaction may trigger the emotional preservation (ususally a hyperactive amygdala and a hypoactive PFC) found in depression~\cite{Siegle}, anxiety~\cite{David} and other fear disorders~\cite{Gehlert}.

\item {\bf Hippocampus -- prefrontal cortex}

Clinical and experimental studies implicate both hippocampus and PFC in several aspects of learning and memory. Not surprisingly, the two units are stongly interconnected and modulate each other's activity in a complex manner.
Hippocampal inervation of the PFC is mainly excitatory and originates fron the temporal CA1/subiculum region and projects to the prelimbic, medial orbital and infralimbic areas~\cite{Jay}. Conversely, hippocampal memory supression is (at least for non-psychiatric populations) under the control of prefrontal regions~\cite{Banich}.

Cerqueira et al.~\cite{Cerq} explain how stress can influence the integrity of the hippocampus-PFC pathway, and thereby explain some of the neurobiological deficits triggered by stress that can't be attributed to hippocampal lesions. The study correlated stress exposure with an observed volumetric reduction in the upper layers of the mPFC which could not be accounted for by neural loss, but rather by dendritic atrophy and retraction of the pyramidal neurons in layers II and III of the mPFC (also see ~\cite{Cerq2007}). Although the hippocampus-mPFC pathway was shown to be impaired even by a single episode of acute stress~\cite{Rocher}, this stress-induced atrophy seems to be reversible~\cite{Radley}.

\item {\bf Amygdala -- hippocampus} The amygdala impact on the hyppocampus is best represented not by neural pathways, but by the indirect autonomic and endocrine effects initiated in the amygdala in response to stress, which lead to hippocampus impairment and functional reduction (as described above). Conversly however, studies such as~\cite{Maren} have opened the possibility that hippocampal projections to the B might be important for contextual contributions in fear extinction.

\end{enumerate}

\begin{figure}[!h]
\begin{center}
\includegraphics[scale=0.15]{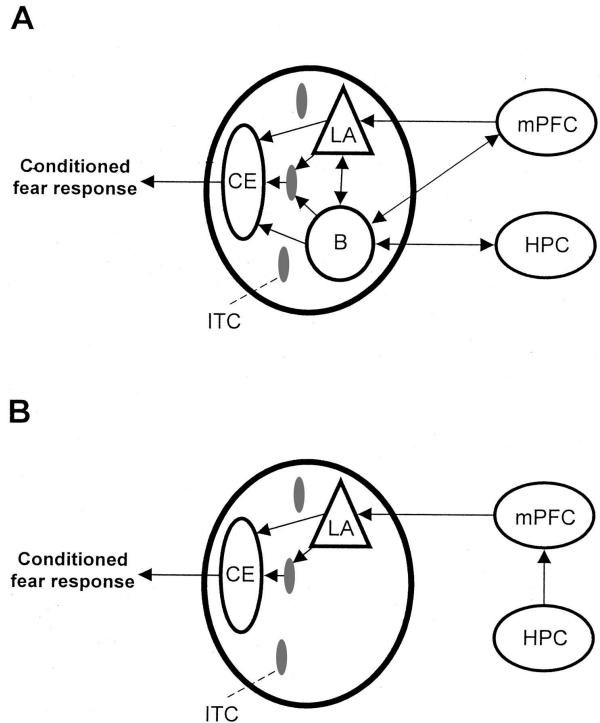}
\caption{\emph{Two schematic models of the amygdala-hippocampus-mPFC interactions, as illustrated by Sotres-Bayon et al.}~\cite{SB}.}
\end{center}
\end{figure}

\section{The mathematical model}

 Our theoretical, simplified model quantifies the amygdala-mPFC mutual regulation in a way which can be studied and understood analytically. The pathways between the two regions, as well as the self-modulation within each region are represented by linear terms, while the indirect hippocampus-modulated influences are expressed as nonlinearities. As we don't consider bilaterality, our model is constructed as a 2-dimensional dynamical system in which the variables $a=a(t)$ and $p=p(t)$ are levels of activation of amygdala and prefrontal cortex, respectively. (In a data-driven model, one may think of these variables as BOLD percent signal change.)

The strengths of the interactions are tuned differently for different individuals. In our model, this tuning is quantified by a set of parameters, so that the time-evolution of the system, and ultimately its asymptotic behavior, depend on the choice of these parameter values. While making no claim to illustrate exactly the complex fear reaction, the model should rather be seen as a metafor of the brain undergoing stress, supporting the limbic dysregulation hypothesis. This has no imediate practical value. However, if explored further, the idea may turn out to be clinically priceless, as it suggests ways in which a more quantitative approach would be helpful to the field of psychiatry. New clinical paradigms, could be developped to test and use this hypothesis, as we will later explain. 

In the next sections, we will show how for different values of the parameters, the system can exhibit dramatically different time evolutions: some corresponding to normal physiology and behavior, some corresponding to typical symptoms of schizophrenia.

\subsection{A linear model. Pros and cons.}

Let's start with the strong (and, as we will see, unreasonable)
assumption that the amygdala-mPFC dynamic is linear, and criticize
the flows inherent to this oversimplification. One would not expect, in
general, for complex phenomena in the brain to behave in a linear
fashion. It is a well-known mathematical fact that linear systems don't exhibit any ``interesting" behavior. As a rule of thumb an isolated fixed point is either a global attractor or repeller, a center or a saddle point (depending on the nature of the Jacobian's eigenvalues at that
point). Moreover, linear systems have no limit cycles,
due to the Poincare-Bendixon theorem. The only
circumstance in which cycles exist is if the system has a
``center'', i.e. a fixed point surrounded by accumulating cycles,
neither attracting nor repelling them. (Mathematically, centers
happen when the Jacobian has two conjugate imaginary roots.)

In this context, if the parameters of the system change, the
behavior of a fixed point may change from being a global attractor
to being a repeller, going through a transitional stage
(\emph{bifurcation}). The bifurcation corresponds to a \emph{critical} 
set of the parameters for which the fixed point is a center,
and thus represents a point of sudden transition in the
dynamics (see Figure 1). As one would want to believe that brain processes have somewhat more fail-safe mechanisms, this is a feature of the linear
model that we will try to correct later.

As an illustration, we construct a very simple and plausible linear
model for the basic interaction between the amygdala and mPFC. This
toy version of the model will leave out the autonomic and endocrine
feedback loops, and will therefore neglect their influence on the
hippocampus function. All other region interactions will be assumed linear.

\begin{align*}
\dot{a} &= -\mu_1 a - k_1 p+I-\gamma_1 H \\
\dot{p} &= k_2 a + \mu_2 p +\gamma_2 H
\end{align*}

\noindent where $I,H,\mu, k_1, k_2, \gamma_1, \gamma_2 >0$.

The amygdala activation $a$ is driven by four terms: the input
$I>0$ (corresponding to the background environmental stimuli), the self-inhibition $-\mu_1 a$  (the amygdala ``resilience to stress") and the prefrontal and hippocampal modulations $-k_1 p$ and $\gamma_1H$ (the hippocampus is assumed to have constant activation $H$, since it receives no feedback). In the absence of the PFC term, the asymptotic equilibrium for $a$ would be $a=\displaystyle{\frac{I-\gamma_1 H}{\mu_1}}$. The PFC activation $p$ is driven by: the amygdala excitatory input $k_2 a$, a hippocampal excitatory modulation $\gamma_2 H$ and a self-excitation $\mu_2 p$.

\begin{figure}[!h]
\begin{center}
\includegraphics[scale=0.35]{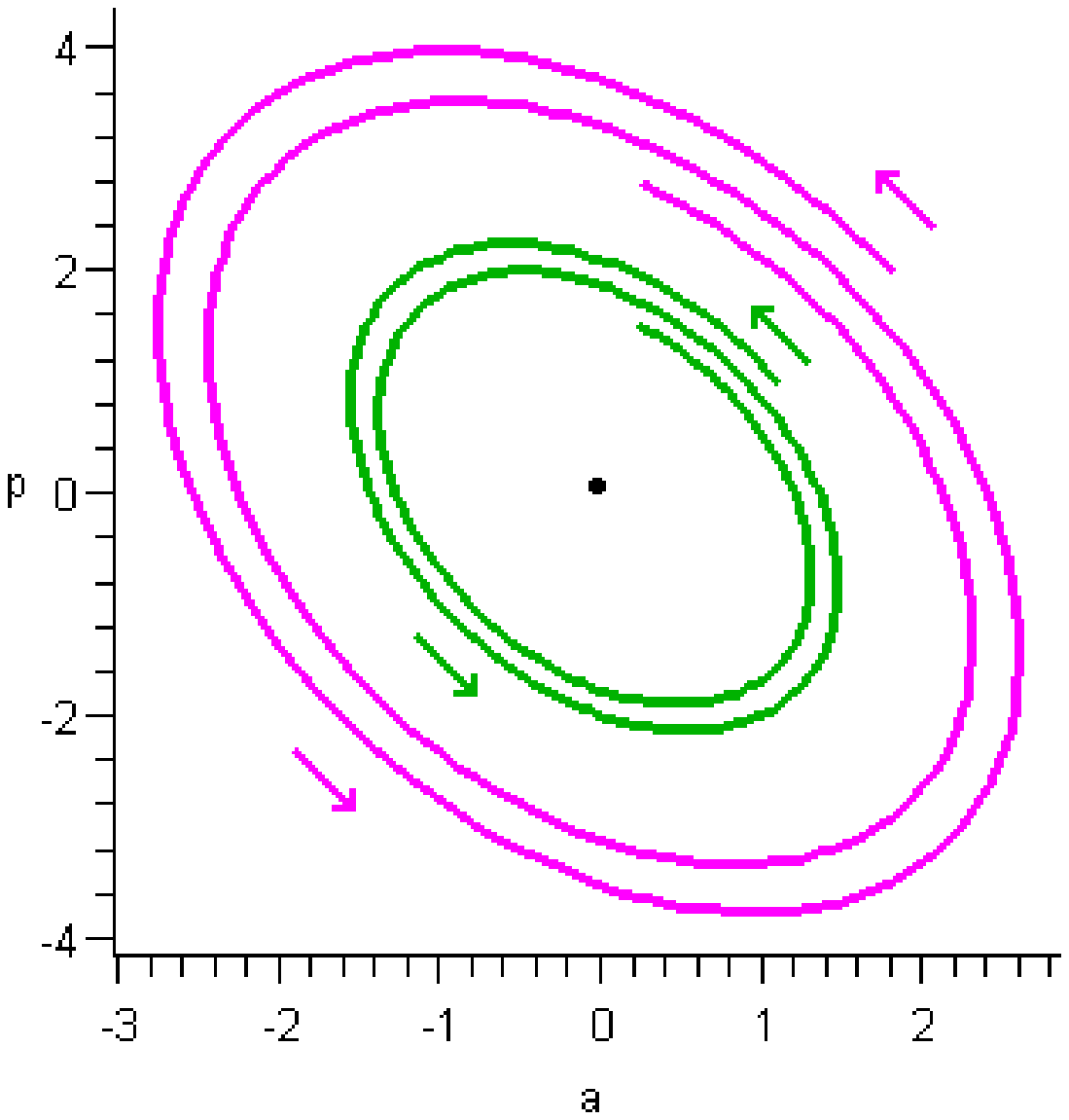}
\includegraphics[scale=0.35]{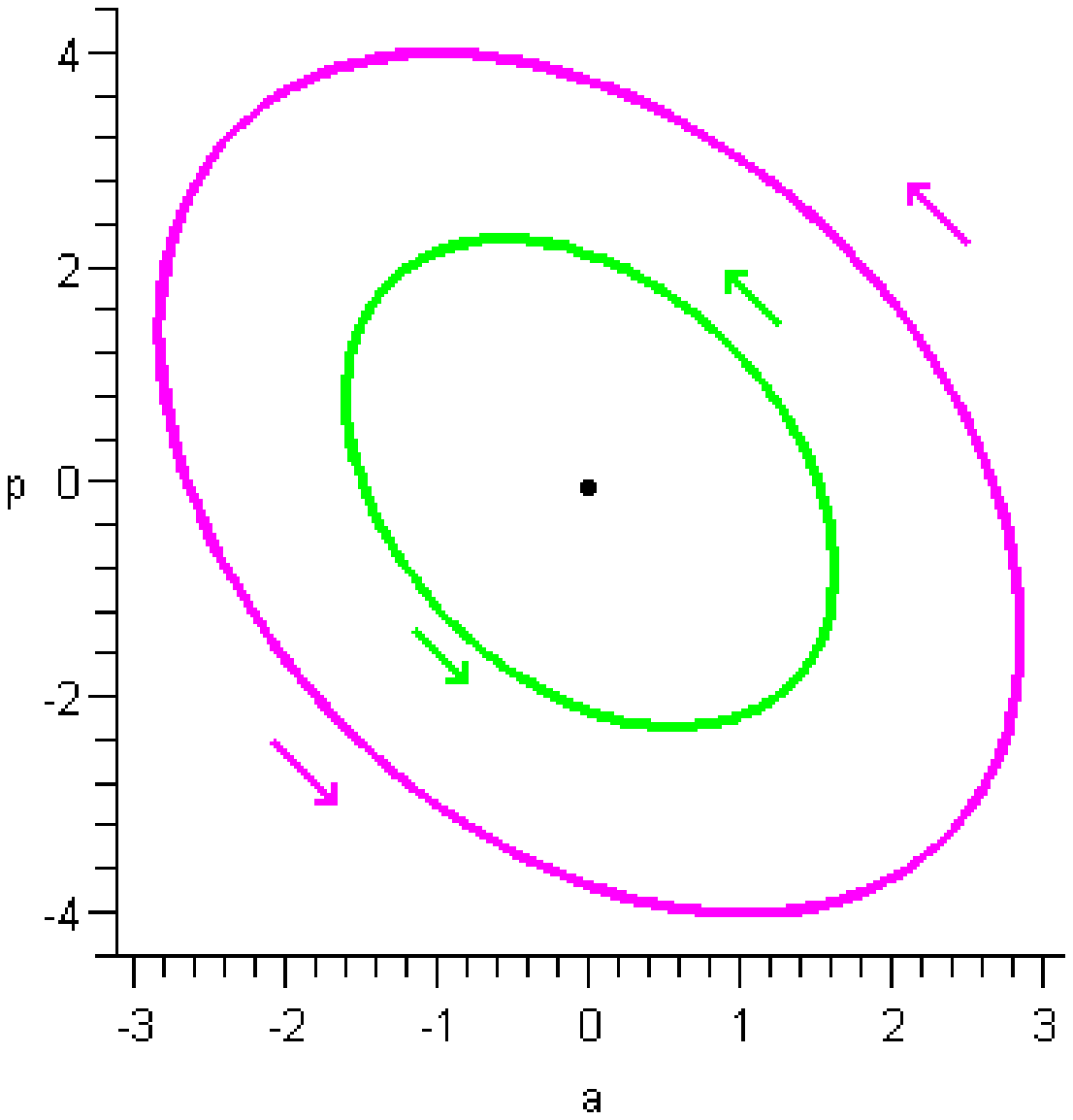}
\includegraphics[scale=0.35]{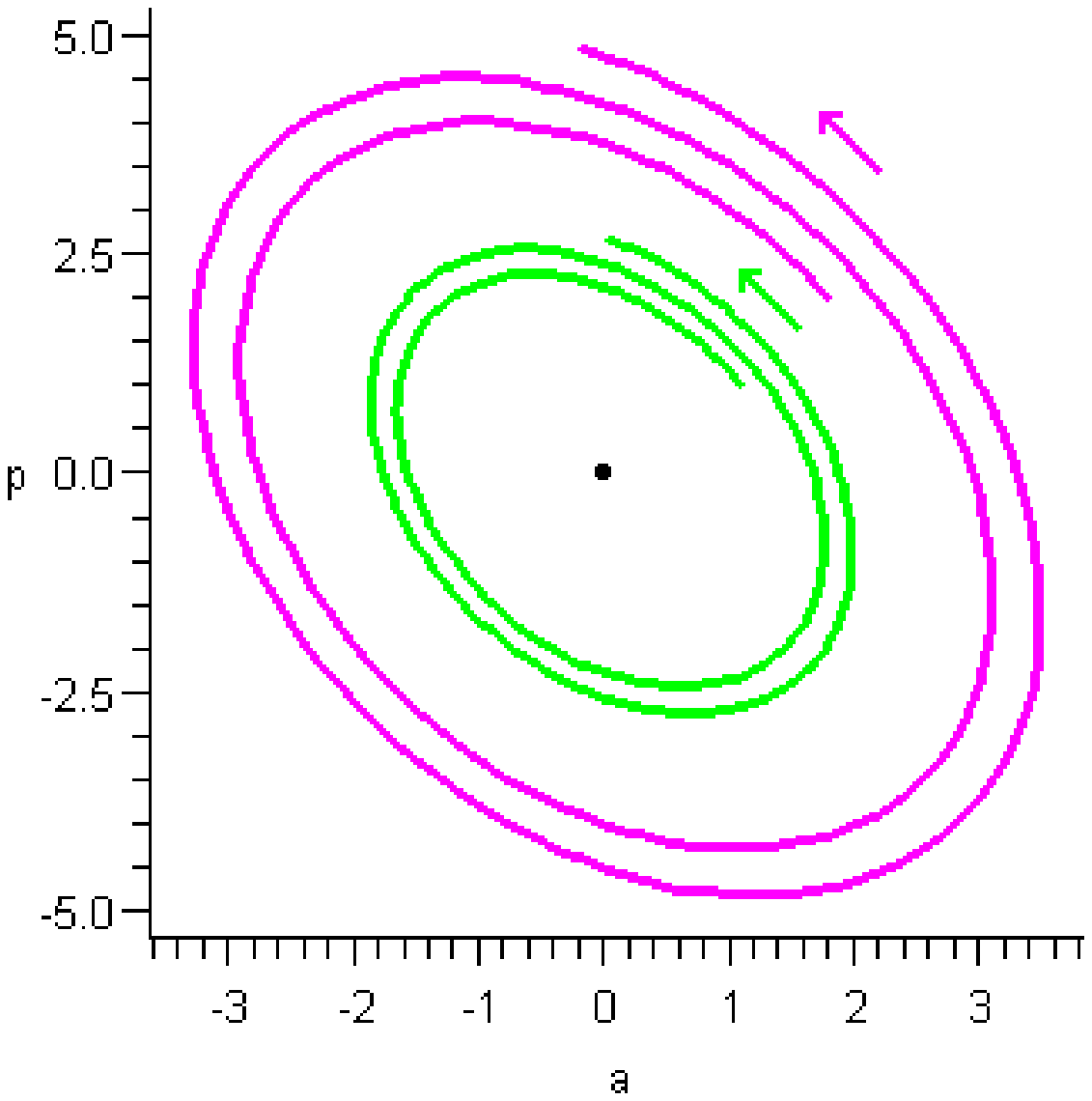}
\caption{\emph{Phase portraits of the linear model for $\mu_2=1, k_1=2, k_2=4$ and different values of the parameter $\mu_1$ The fixed point $(a^*,p^*)$ is a global attractor for $\mu_1=1.1$ (left), a center for $\mu_1=1$ (middle) and a repeller for $\mu_1=0.9$.}}
\end{center}
\end{figure}

In a ``brain'' with a well-balanced feedback, this should euristically work as follows: If the amygdala activation level is high, its excitatory
effect on the mPFC will drive $p$ up, which in turn will inhibit $a$
and make the system converge towards an equilibrium. In a ``brain''
with an overactive amygdala (low $\mu_1$ value), convergence fails to
happen.

The Jacobian matrix:

$$D=\left( \begin{array}{cc}
\mbox{$-\mu_1$}&\mbox{$-k_1$}\\
\mbox{$k_2$}&\mbox{$\mu_2$}
\end{array} \right)$$

\noindent has determinant $\Delta=k_1 k_2 - \mu_1 \mu_2$. We will
assume that $\Delta=k_1 k_2 - \mu_1 \mu_2>0$, i.e. we work in a regime where 
the interconnections prevail over self-modulations in the two regions.

Then the system has always a fixed point (equilibrium) at

$$(a^*,p^*)=\left( \frac{k_2 I+H(\mu_1 \gamma_2-\gamma_1 k_2)}{\Delta}, \frac{-\mu_1 I+H(\gamma_1 \mu_2-k_1 \gamma_2)}{\Delta}
\right)$$

The equilibrium depends on the outside stress level and on the hippocampus
activation, as well as on the mutual and self excitation and
inhibition parameters. The fixed point is globally attracting if $\mu_1 > \mu_2$, i.e. if the amygdala self-inhibition is strong enough to exceed the PFC self-excitation. However, as $\mu_1$ decreases, the stability of the fixed point changes, and for $\mu_1<\mu_2$ it becomes a global repeller. A system operating under such values of the parameters ($\mu_1<\mu_2$) will have all trajectories pushed away from the repeller and never stabilize. It is hard to believe that the brain operates with switch between stable and unstable behavior so suddenly. 

So where could the nonlinearity come from and what does it do?

\subsection{The nonlinear model}

Nonlinearity in a two-dimensional system introduces an interesting
characteristic feature: possible existence of limit-cycles. Consider the
general form of a 2-dimensional nonlinear system with a fixed point
at the origin:

$$\left( \begin{array}{c} \dot{x} \\ \dot{y} \end{array} \right) =
D \left( \begin{array}{c} x \\ y \end{array} \right) + N(x,y)= \left(
\begin{array}{cc} d_{11} x + d_{12} y + N_1(x,y) \\ d_{21} x + d_{22} y
+ N_2(x,y) \end{array}\right)$$

\noindent where $D$ is the Jacobian matrix and $N$ is the nonlinear part, both parameters-dependent.

The eigenvalues $\lambda_1$ and $\lambda_2$ of $D$ may take real or
complex values that satisfy $\lambda_1+\lambda_2=\text{trace}(D)=\tau$
and $\lambda_1 \lambda_2 =\Delta=\det(D)$. In general, if $\Delta>0$, than the
origin is either an attracting node or spiral (in case $\tau<0$) or
a repelling node or spiral (in case $\tau>0$). At the parameter values where
$\tau=\lambda_1+\lambda_2=0$, the system exhibits a bifurcation. In
particular, if $\lambda_1$ and $\lambda_2=\overline{\lambda_1}$ are imaginary, then we
may have a \emph{Hopf bifurcation}. The way the local dynamics of the
system changes at a Hopf bifurcation is described by the Lyapunov
number, which depends on both the linear part $D$ and the nonlinear
part $N$, as described below~\cite{Perko}. If $N_1, N_2 \colon
\mathbb{R}^2 \to \mathbb{R}$ are analytical, with expansions:

$$N_1(x,y)=\sum_{i+j \geq 2}{a_{ij}x^iy^j}=(a_{20}x^2+a_{11}xy+a_{02}y^2)+(a_{30}x^3+a_{21}x^2y+b_{12}xy^2+b_{03}y^3)+...$$

$$N_2(x,y)=\sum_{i+j \geq 2}{b_{ij}x^iy^j}=(b_{20}x^2+b_{11}xy+b_{02}y^2)+(b_{30}x^3+b_{21}x^2y+b_{12}xy^2+b_{03}y^3)+...$$

\noindent then the Lyapunov number:

\begin{eqnarray*}
\sigma &=& \frac{-3 \pi}{2 d_{12} \Delta ^{3/2}} \{ d_{11}d_{21}(a_{11}^2+a_{11}b_{02}+a_{02}b_{11})+D_{11}d_{12}(b_{11}^2+a_{20}b_{11}+a_{11}b_{02})\\
&+& d_{21}^2(a_{11}a_{02}+2a_{02}b_{02})-2d_{11}d_{21}(b_{02}^2-a_{20}a_{02})-2d_{11}d_{12}(a_{20}^2-b_{20}b_{02})\\
&-& d_{12}^2(2a_{20}b_{20}+b_{11}b_{20})+(d_{12}d_{21}-2d_{11}^2)(b_{11}b_{02}-a_{11}a_{20})\\
&-& (d_{11}^2+d_{12}d_{21})[ 3(d_{21}b_{03}-d_{12}a_{30})+2d_{11}(a_{21}+b_{12})+(d_{21}a_{12}-d_{12}b_{21}) ] \}
\end{eqnarray*}

If $\sigma \neq 0$, a Hopf bifurcation occurs at the critical
value of the parameters where $\tau=0$. More precisely:\\

(1) If $\sigma<0$, the origin is attracting for $\tau \leq 0$, and
the system has no limit cycle. For $\tau>0$, the origin becomes
repelling, but a circular stable limit cycle forms around it, whose
radius increases with $\tau$.

\begin{figure}[!h]
\begin{center}
\includegraphics[scale=0.35]{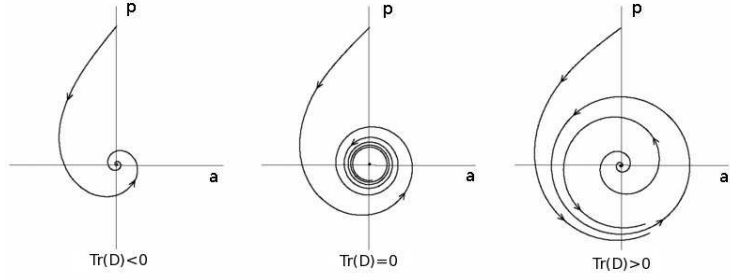}
\caption{\emph{When $\sigma<0$, the system exhibits a supercritical Hopf bifurcation.}}
\end{center}
\end{figure}

(2) If $\sigma>0$, the system has a unique unstable limit cycle that
surrounds the stable origin for $\tau<0$. The radius of the cycle
decreases with $\tau$. At $\tau=0$, the unstable cycle collapses
into the origin, making it unstable for $\tau>0$.

\begin{figure}[!h]
\begin{center}
\includegraphics[scale=0.35]{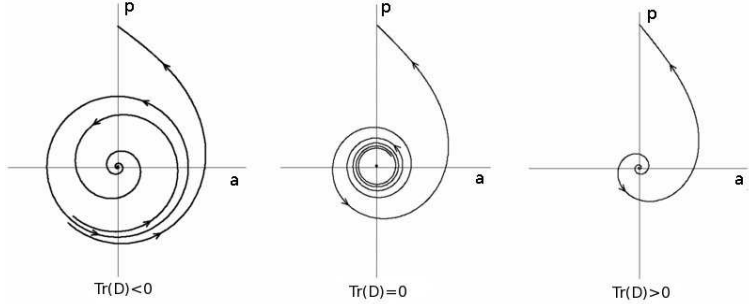}
\caption{\emph{When $\sigma>0$, the system exhibits a subcritical Hopf bifurcation.}}
\end{center}
\end{figure}

The change in dynamics is still very sudden at a Hopf
bifurcation (even though the parameters change smoothly), but has much
more subtle implications. To interpret this phenomenon in our clinical context, let's fist construct as before a more particular example of a nonlinear model, as an extension of our linear toy-model.\\

Staying faithful to our original assumption, we include the
nonlinearities as the contribution of the autonomic and endocrine
response to stress. As mentioned in the introduction, experimental
studies showed that high levels of cortisol due to the amygdala
stress-reaction have a detrimental effect on the PFC, in
two known distinct ways: its neurotoxic effects on hippocampal cells
and its suppression of synaptic function between hippocampus and the receptive PFC areas. The new system, including these nonlinear contributions, will take the form:

\begin{align*}
\dot{a} &=-\mu_1 a - k_1 p+I-\gamma_1(H-f(a))\\
\dot{p} &=k_2 a - \mu_2 p+\gamma_2(H-f(a)+g(a,p))
\end{align*}

\noindent Here the term $-f(a)$ signifies the structurally detrimental effect that amygdala overactivity and the subsequent hypercortisolemia has directly on the hippocampus. The term $g(a,p)$ refers to the synaptic remodeling induced by high cortisol on the PFC afferents from the hippocampus, and hence it only affects $p$. 

Under the change of variables $x=a-a^*$ and $y=p-p^*$, the system becomes:

\begin{align*}
\dot{x} &=\mu_1 x - k_1 y+\gamma_1 h(x)\\
\dot{y} &=k_2 x -\mu_2 y-\gamma_2 (h(x)-j(x,y))
\end{align*}

\noindent where $h(x)=f(a)=f(x+a^*)$ and $j(x,y)=g(a,p)=g(x+a^*,y+p^*)$.

\noindent To fix our ideas, we will take both terms to be of the simplest possible nonlinear forms: $h(x)=\gamma x^2$ to reflect the decease in $H$ activation and $j(x,y)=\delta xy$, for the suppression of Hebbian-like synapses in mPFC. Recall that, in our original system, this corresponds to $f(a)=\gamma (a-a^*)^2$ and $g(a,p)=\delta (a-a^*)(p-p^*)$.

 The system has a fixed point at $(x^*,y^*)=(0,0)$, and exhibits a Hopf bifurcation at $\mu_1=\mu_2$, with Lyapunov number

$$\sigma=\frac{3 \pi}{2 \Delta^{3/2}}(\delta \gamma_2 + 2 \gamma \gamma_1) \left[ \mu_2(\delta \gamma_2 - \gamma \gamma_1) + k_1 \gamma_2 \gamma \right]$$

Applying the theoretical results we discussed above, we conclude that: if $\displaystyle{\frac{\delta}{\gamma}+\frac{k_1}{\mu_1}<\frac{\gamma_1}{\gamma_2}}$ (i.e., $\sigma<0$), then the system has a supercritical Hopf bifurcation at $\mu_1=\mu_2$, and if $\displaystyle{\frac{\delta}{\gamma}+\frac{k_1}{\mu_1}>\frac{\gamma_1}{\gamma_2}}$ (i.e., $\sigma>0$) then the system has a subcritical Hopf bifurcation at $\mu_1=\mu_2$.

\section{Discussion}

This theoretical model is more a ``phylosophical" than ``physiological" illustration of the limbic dysregulation and neurotoxicity hypotheses. It presents the working brain in a light that permits interpretation of its ``stress vulnerability" and ``hippocampus deficit" as parameters that vary continuously, determining its regulation and function. The focus of this interpretation is on the idea that, although these parameters change smoothly over a whole continuum of possible values, there are critical/threshold values, which, when passed, could suddenly and completely change the system's dynamics.

The two parameters on which we focus, whose tuning determines the behavior of the system, are the amygdala sensitivity to stress, represented by $\mu_1$ and the more diffuse hippocampal/prefrontal vulnerability to cortisol neurotoxicity, represented by the Lyapunov number $\sigma$. Indeed, staying faithful to the stress/vulnerability hypothesis of schizophrenia, our interpretation will regard $\sigma$ as the ``disease-quantifying'' parameter: negative values of $\sigma$ correspond to normal limbic 
regulation, while positive values of $\sigma$ quantify risk for developing schizophrenia, and, in more advanced stages, severity of the illness. On the other hand, larger values of $\mu_1$ correspond to a more stress-resilient amygdala, while smaller values of $\mu_1$ signify a more stress-reactive amygdala. We could think of this parameter as quantifying the amygdala responsiveness to stress, which in literature has been related to mental conditions such as depression, or anxiety disorders~\cite{SB}, but which is not the signature of schizophrenia. We can verify whether this paradigm is clinically plausible by testing what happens if we apply a brief stress increase to the system (which in real life may come in the form of a taumatic event). We quantify the burst of stress by boosting the amygdala to a high initial state. We observe wheter the system returns to homeostasis by checking if the respective trajectory eventually stabilizes.

Suppose $\sigma<0$. For high amygdala resilience $\mu_1>\mu_2$, all trajectories converge to the global attractor, so the initial condition is irrelevant: the time evolutions stabilize after any stimulus, if sufficient time is allowed to pass. For $\mu_1<\mu_2$, the situation is changed by the formation of an attracting limit cycle(Figure 4a). After a short stress burst, the duo amygdala-PFC slowly stabilizes towards the cycle. Note that, although dampened in time, the system continues to oscillate in both cases. We will return to this idea later. The memorable feature of the $\sigma<0$ regime is that, although stability of the fixed point changes at the bifurcation $\mu_1=\mu_2$, the role of the attractor is assumed by a limit cycle. The fact that the amygdala-PFC pair exhibits in some people wider oscillations that don't seem to dampen in time could be a mark of low amydgala self-inhibition.

\begin{figure}[!h]
\begin{center}
\includegraphics[scale=0.45]{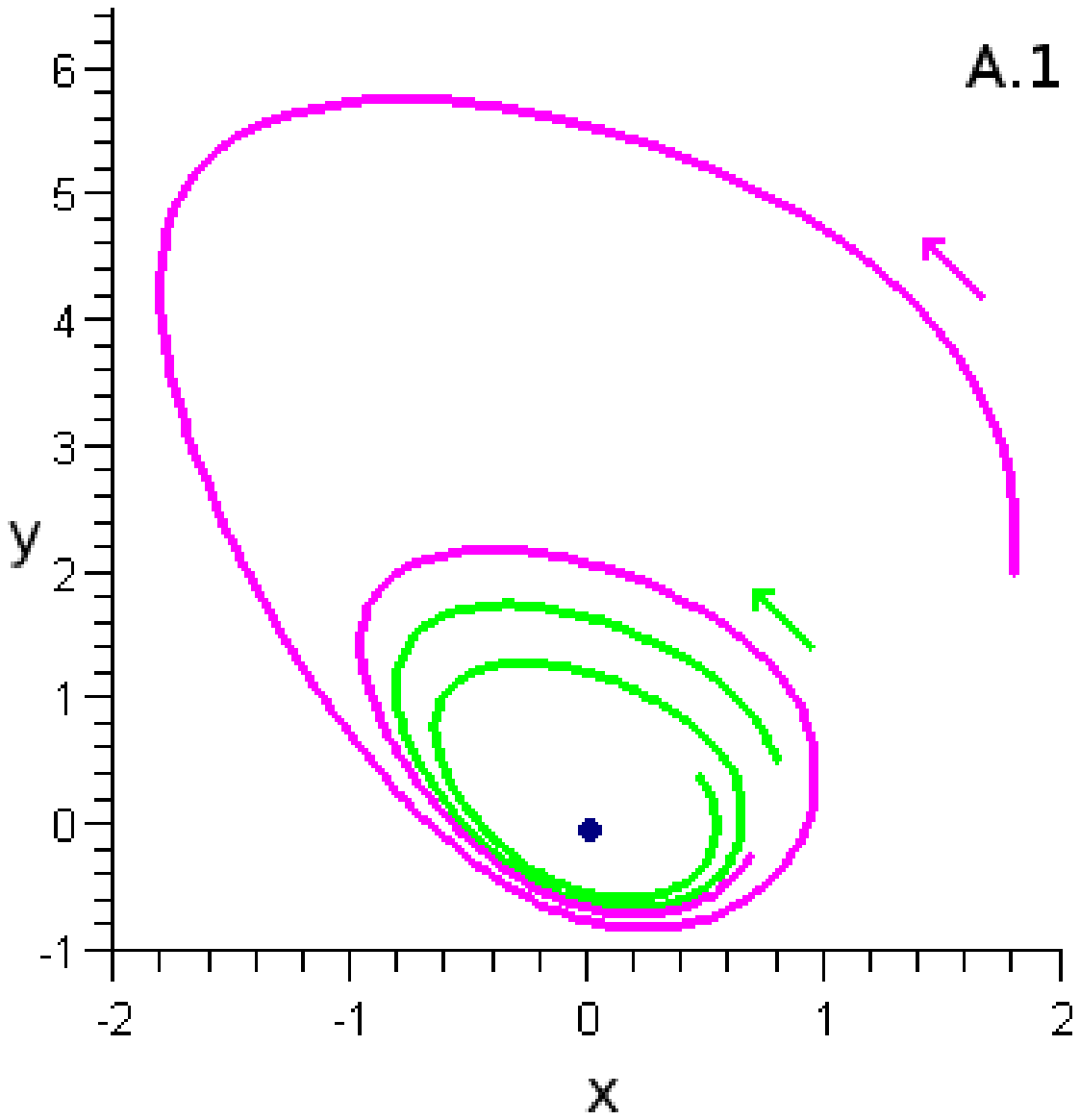}
\includegraphics[scale=0.45]{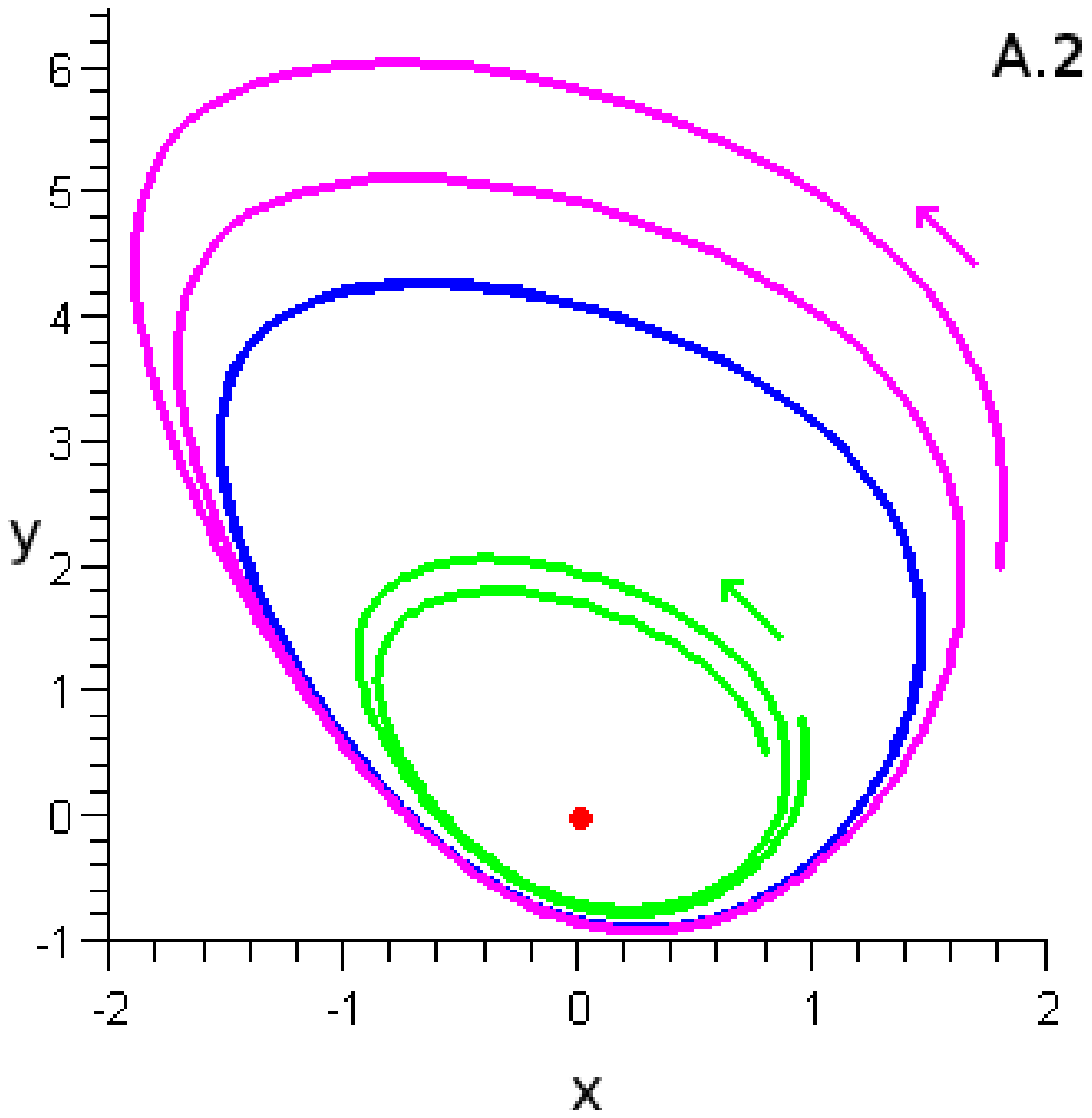}\\
\includegraphics[scale=0.45]{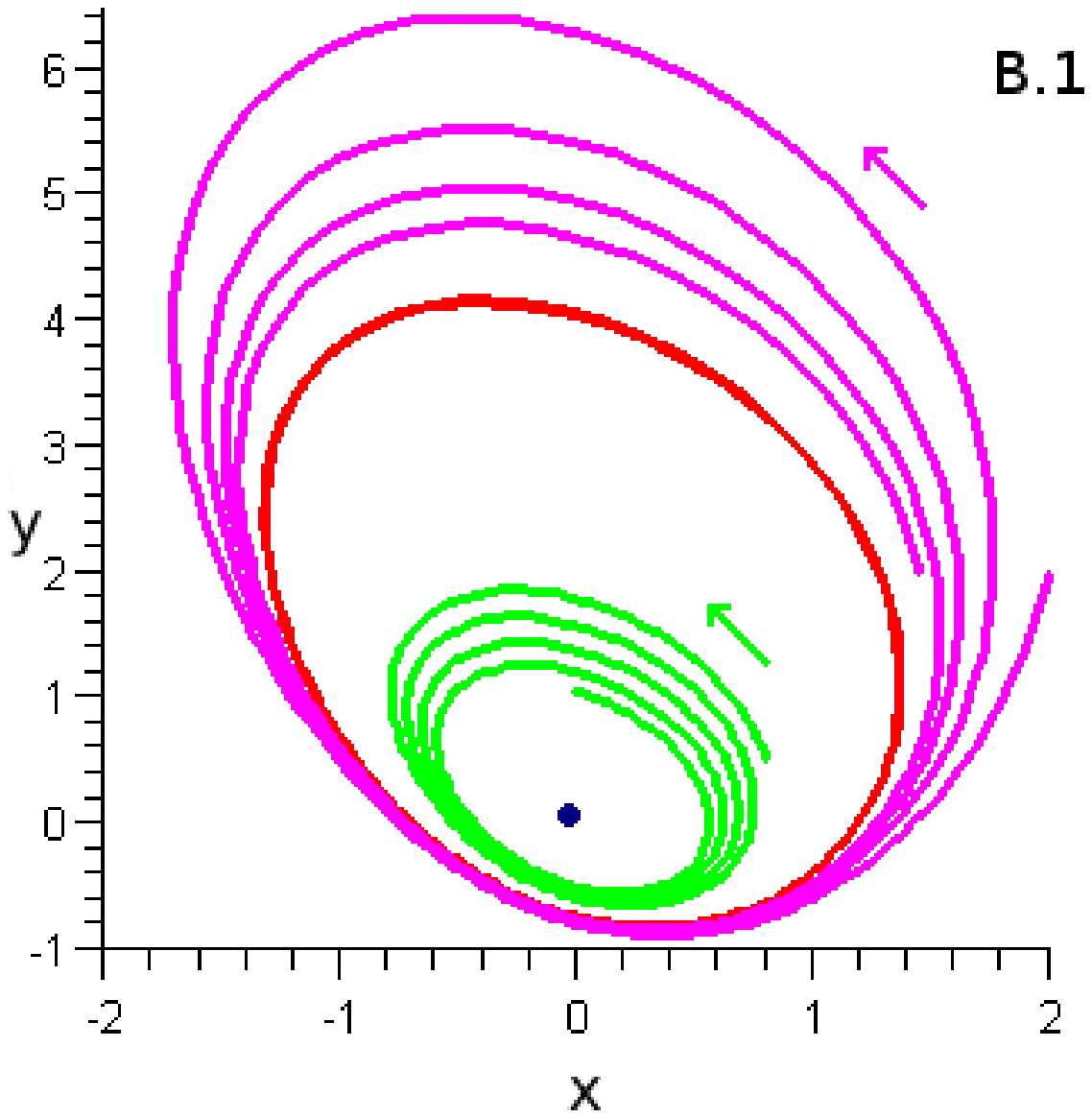}
\includegraphics[scale=0.45]{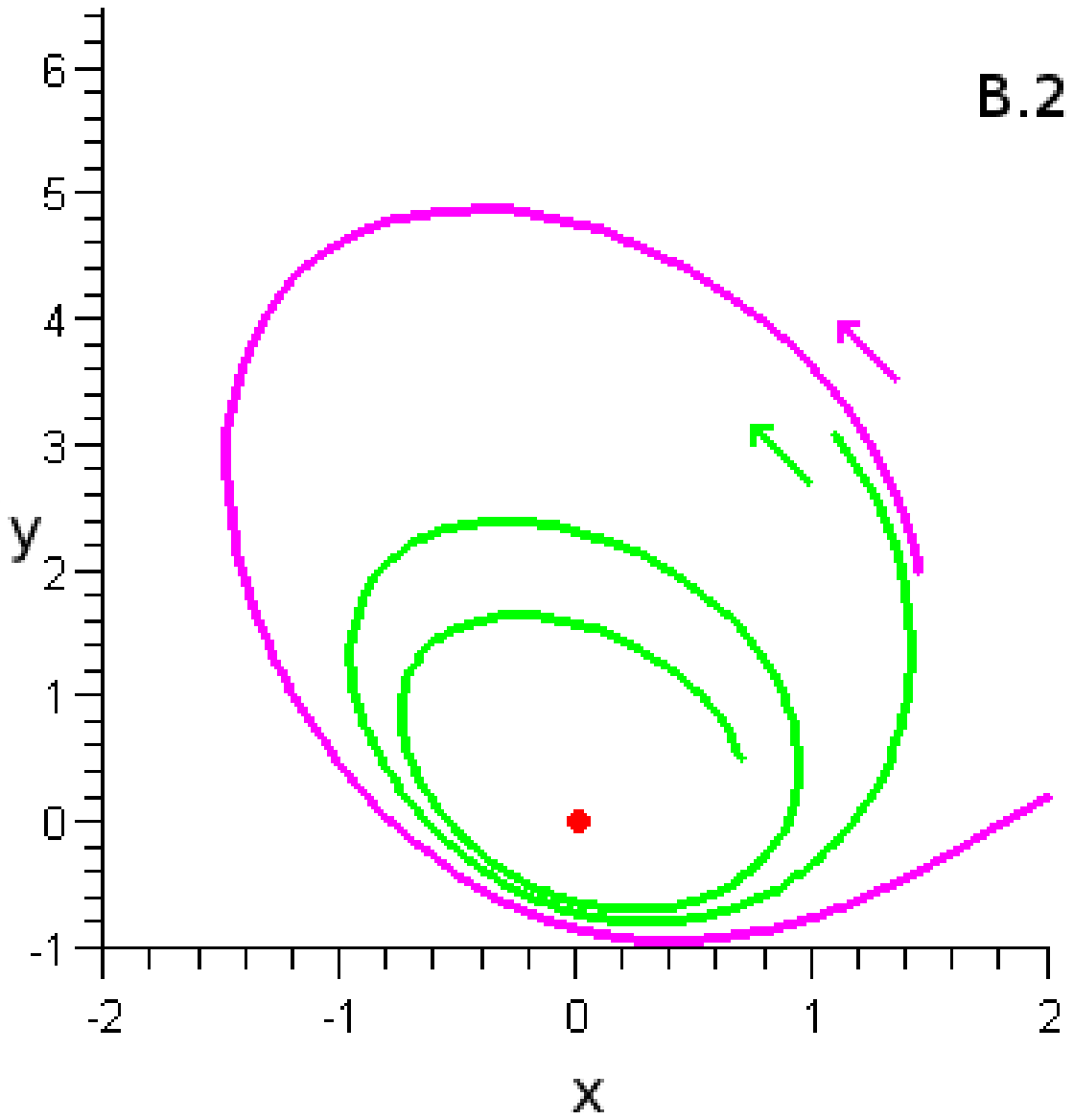}
\caption{\emph{We illustrate the dynamics of the nonlinear model for a particular example of parameter values. Fixed parameters: $\mu_2=1, k_1=2, k_2=4, \gamma_1=2, \gamma_2=0.5, \gamma=1$. A. The system exhibits a supercritical Hopf bifurcation for $\delta=1$. The phase portraits show the local dynamics around the origin for $\mu_1=0.9$ (panel A.1) and $\mu_1=1.1$ (panel A.2). As $\mu_1$ crosses the critical value $\mu_2=1$, the attracting origin becomes repelling and surrounded by an attracting limit-cycle. B. The system exhibits a subcritical Hopf bifurcation for $\delta=3$. As $\mu_1$ crosses the critical value $\mu_1=1$, the repelling cycle surrounding the attracting origin ($\mu_1=0.9$, shown in B.1) collapses into the origin and changes it to a global repeller ($\mu_1=1.1$, shown in B.2).}}
\end{center}
\end{figure}

When $\sigma >0$, the situation changes completely. In the regime $\mu_1>\mu_2$, the system has a locally attracting fixed point, surrounded by a repelling cycle, whose radius gets smaller with smaller $\mu_1$ and with larger $\sigma$. The basin of attraction of the stable point is the interior of the cycle: any initial state inside this basin will converge in time towards the fixed point, and any initial state outside will spiral out to infinity. When $\mu_1=\mu_2$, the cycle disappears, so when $\mu_1<\mu_2$ the fixed point is globally repelling (Figure 5b). The behavior of the model in the positive $\sigma$ regime is representative for schizophrenic dysregulation. A stimulus may elevate amygdala to a value which places the corresponding state outside of the attraction basin, preventing convergence from ever happening. If alowed to follow its natural evolution, the trajectory would perform larger and larger oscillations, corresponding to the cyclic psychotic behavior observed in patients. At some point during this evolution, the patient may enter clinical treatment; antipsychotic medication may succeed to temporarily alter this time-course. It is interesting to note that, as we increase the value of $\sigma$, the attraction basin shrinks around the attracting point, so that even weak stressful stimuli may push the state outside the attraction basin. This makes the Lyapunov number $\sigma$ a good quantifier of the risk and severity of the disease: the larger $\sigma$, the more likely it is for the time-evolution to be thrown outside of the convergence range even by small perturbations. This relates to the fact that highly vulnerable individuals may develop psychotic behavior even after common daily stress that may appear benign to others.    

We want to conclude the analysis of the phenomena described above with a few time-frame considerations. Our model works in parallel with behavioral research that relates stress with first outbreak~\cite{Hazlett} and relapse~\cite{Ventura}. The time-frame that the model addresses is thereby of the order of days, even weeks, allowing possible psychotic behavior to develop after a stressful event. Cycles and oscillations in brain activity, as well as regulation and return to homeostasis have been already the subject of a wide variety of studies, addressing very different time scales. Elecrophysiology studies revealed high frequency oscillations in neuronal brain activity, and correlated synchronization of theta rhythms in the amygdalo-hippocampal pathways with retrieval of conditioned fear~\cite{Pare1}~\cite{Pare2}. It is also known that that the brain has a circadian rhythm, so that its activity oscillates according to a daily pattern~\cite{Guilding}. Additionally, recent imaging studies focus on the dynamics of certain regions of interest in response to visual stimuli, such as facial expressions. (See Figure 6, which illustrates comparatively the behavior of the amygdala and a prefrontal area, in control subjects and schizophrenic patients.) Although our model does not address such short-scale phenomena, these findings better describe its place as just one level in a general complex picture.

\begin{figure}[!h]
\begin{center}
\includegraphics[scale=0.4]{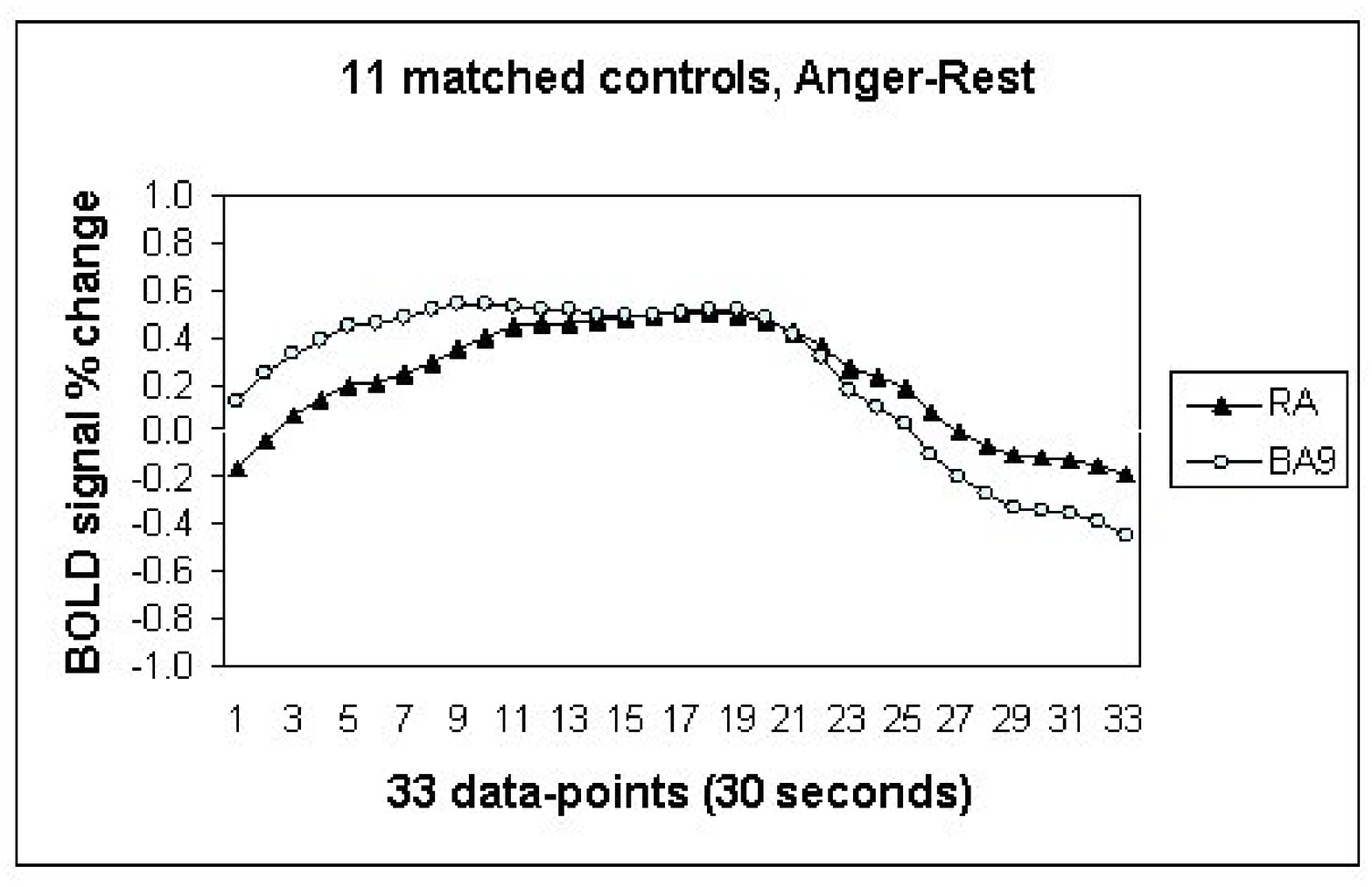}
\includegraphics[scale=0.4]{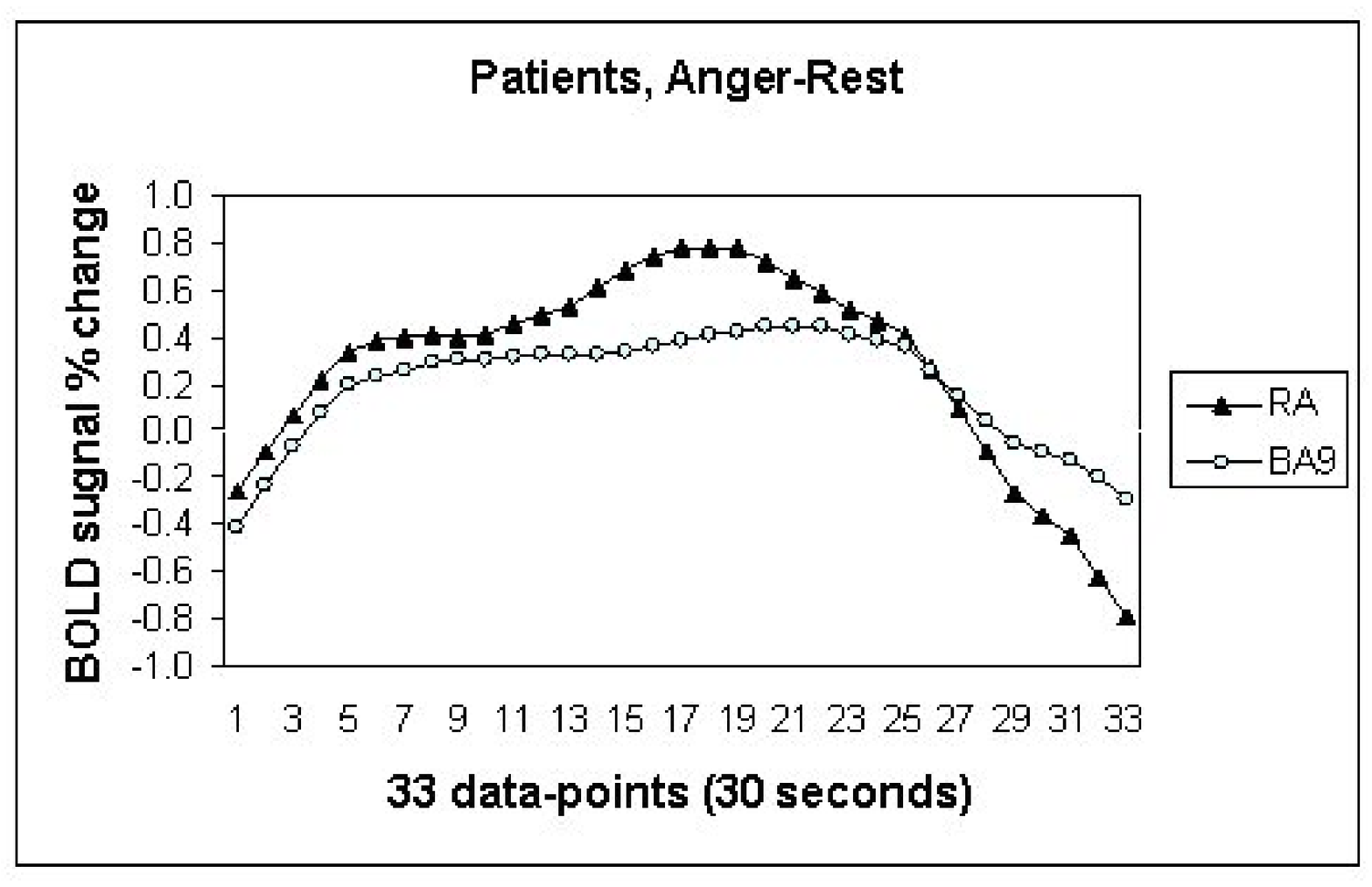}
\caption{\emph{A. fMRI data on amygdala and BA45 activation under visual stimulation with Pictures of Facial Affect. The data has been averaged over all stimulation blocks, as well as over the control and population samples, respectively (samples sizes $N=11$)}~\cite{Anca}.}
\end{center}
\end{figure}

Further testing of the model should be willing to compromise precision for duration. Firstly, the design does not need to focus on identifying the exact brain areas involved in the process, but rather in estimating the average activation in generic areas such as amygdala, hippocampus and mPFC. Secondly, these activations need to be measured over an extended time-period, allowing the feed-back cycle to repeatedly loop around, so that the long-term dynamics could become apparent. In this context, EEG readings would be ideal, as they can be taken at length, without drastically restricting the subject's normal routine. To create the required set-up and take the measures following a stress burst, the testing should be ideally planned for immediately after a scheduled stressful event such as an exam, or first time ski-diving (see $http://www.bme.sunysb.edu/people/lstrey/research.htm\#2$). Another variable of interest that can be very conveniently measured during  a long-term routine is salivary cortisol, which is believed to reflect stress-related hypecortisolemia. In our model this variable is one of the factors determining the system's nonlinear behavior, so observing its evolution would help us understand these nonlinearities better. To the best of my knowledge, these testing paradigms are -- if not already being used in other tests -- then fairly easy to design and reproduce.

\section{Conclusions}

A natural question to addresses is the significance of this model. Theoretically, the model shows how two different clinical systems (with very similar underlying rules, and only slightly different parameter values) can exhibit drastically different long term behavior if started under the same initial conditions. The literature talks about the ``continuum" of human behavior and the practical difficulties of establishing a normality/pathology threshold. Such a bifurcation could constitute the needed threshold for clinical evaluations. Practically, if the model proves to be valid, both diagnosis of illness and quantification of its severity can be achieved by calculating the Lyapunov number of a system constructed from clinical measures. As computer capabilities have sky-rocketed over the past few decades, fast and accurate algorithms~\cite{Geist}~\cite{Chon} have become available to compute such system invariants from time-series.

On a slightly different note, our model supports the idea that the dynamics of a diseased system is not driven randomly, but is rather only apparently random due to its complicated behavior over short time periods. This idea is very important for clinical treatment, as it suggests that the deterministic behavior of a system can be changed by proper tuning of the parameters. In general, the drugs currently in use only temporarily alter the time-course of the illness, which typically relapses when interrupting medication. It is possible that medication options could be improved by exploring how drugs can change the parameter values to permanently alter the system and its long-term behavior.

A more sophisticated, higher-dimensional model, although preferable and possible, is a different and much harder task to undertake. As discussed before, the exact physiological underpinnings of the brain interactions are not yet fully known from experimental studies, making an educated choice of parameters very difficult. Furthermore, even the theoretical study of the evolution of a system's dynamics under perturbations is an ongoing mathematical problem. However, we expect that, as the theoretical and experimental knowledge progress, so will our ability to understand and model such complex phenomena.

\vspace{10mm}

\noindent {\Large{\bf Epilogue}}

\vspace{7mm}

While walking in a dark swampy place one night, I had the distinct
feeling that something was following me, with insane reddish eyes,
ready to leap. The darkness, the rotten smell, the chill in the air,
all made my skin wrinkle in little goose-bumps. A stick cracked
under my foot, and my heart skipped a beat. But I did not run
off, neither did my heart stop. A split second later I skipped back
to the reality of my short stroll in the vicinity of a very
populated area, which never harbored red-eyed beasts. I giggled at
the memory of all the horror Hollywood movies that lived
involuntarily on that brief adrenaline peak. The moment was
gone and the heart was beating all its strokes again. But I will
never know what just happened in that brief moment of panic.
How many more years of research are needed to shed light onto
something as seemingly simple, and when will we be able to
understand why such a casual moment sometimes conquers
people and takes over their lives, keeping them is the constant
terror of some red-eyed beast.

\end{document}